\documentclass[preprint2]{aastex}
\usepackage{graphicx}
\usepackage{color}

\begin{document}

\title{Efficiency of Centrifugal Mechanism in Producing PeV Neutrinos From Active Galactic Nuclei}

\author{Osmanov Z.}
\affil{School of Physics, Free University of Tbilisi, 0183, Tbilisi,
Georgia}

\author{Mahajan S.}
\affil{Institute for Fusion Studies, The University of Texas at
Austin, Austin, TX 78712, USA}

\author{Machabeli G. \& Chkheidze N.}
\affil{Centre for Theoretical Astrophysics, ITP, Ilia State
University, 0162 Tbilisi, Georgia}

\begin{abstract}
A several-step theoretical model is constructed to trace the origin
of ultra high energy (UHE) $[1-2]$PeV neutrinos detected, recently,
by the IceCube collaboration. Protons in the AGN magnetosphere, experiencing different
gravitational centrifugal force, provide free energy for the parametric
excitation of Langmuir waves via a generalized two-stream instability. Landau damping of these waves,   outside the AGN magnetosphere, 
can accelerate protons to ultra high energies. The ultimate source 
for this mechanism, the Langmuir-Landau-Centrifugal-Drive (LLCD), is the gravitational energy of the compact object. The LLCD generated UHE protons provide the essential ingredient  in the creation of   UHE neutrinos  via appropriate hadronic reactions; protons of energy~
$10^{17}$eV can be generated in the plasmas surrounding AGN with
bolometric luminosities of the order of $10^{43}$ergs s$^{-1}$. By
estimating the diffusive energy flux of extragalactic neutrinos in
the energy interval $[1-2]$PeV, we find that an acceptably small
fraction $0.003\%$ of the total bolometric luminosity will suffice
to create the observed fluxes of extragalactic ultra-high energy
neutrinos. 

\end{abstract}


\section{Introduction}

The recent discovery of the  ultra high energy (UHE) extra-solar
neutrinos by the Ice Cube collaboration \cite{pev1} is particularly 
interesting. The neutrino trajectories, in contradistinction to those of the UHE charged
particles, are unaffected by the galactic magnetic field,  and can lead the observer back to the
origin of the emanation.

In the Ice Cube announcement, two events - neutrinos of energies
$1.04$PeV and $1.14$PeV with a high significance of observations are emphasised. An
analysis of the observational data between $2010$ and $2013$ also
shows an event that corresponds to even higher energy of $2$PeV
\cite{pev2}. The IceCube observations are very significant 
not only for studying the origin of the UHE neutrinos but in general for exploring the 
astrophysical origin of cosmic rays.

It is usually assumed that the UHE neutrinos are
generated via interactions of UHE protons, during which,
approximately $4\%$ of the initial proton energy is imparted to the
neutrinos, \cite{murase}:
\begin{equation}
\label{enu}  E_{\nu}\approx0.04E_p\simeq
2PeV\epsilon_{p,17}\frac{2}{1+z},
\end{equation}
where $\epsilon_{p,17} = \epsilon_p/10^{17}eV$ is the normalized
proton energy in the cosmic rest frame and $z$ is the redshift of
the corresponding object.

If the proposed hadronic processes were the source of $[1-2]$PeV neutrinos,
one must demand a supply of  protons with energies of the order of $10^{17}$ eV \cite{murase}.
Such a population of UHE protons, could, indeed, come
from the active galactic nuclei (AGN) \cite{kim}. In face, guided by this
observational correlation, we have recently proposed a plausible
mechanism \cite{zev} and worked out a model for driving protons to ultra high
energies in the AGN vicinity.

In the framework of our alternative model the story of UHE charged particles begins with tapping the gravitational energy to excite Langmuir waves in the highly relativistic plasmas present in the  rotating magnetospheres 
of compact objects.  The new mechanism, called the Langmuir-Landau-Centrifugal Drive
(LLCD), has been described in several papers, and in considerable detail in (Mahajan et al. 2013). 
Applied successfully to  demonstrate extreme particle acceleration in plasmas surrounding the  pulsars (Osmanov et al. 2015; Mahajan et al. 2013; Machabeli et al. 2005) and the AGN \cite{zev}, the LLCD operates in following successive steps:

1) Differential and time dependent centrifugal force,
parametrically, destabilizes Langmuir waves. The pumping of
rotational energy into the electric field is extremely efficient in
the vicinity of the light cylinder (LC) surface (a hypothetical
area, where the linear velocity of rotation exactly equals the speed
of light).

2) These growing Langmuir waves, sustained in the bulk plasma (either
electron-positron or electron- proton), damp on a  faster component
of particles, accelerating them further. In the AGN this energy
transfer is further boosted by a Langmuir collapse prior to Landau damping.

3) The overall process is most efficient when the rate of transfer
from rotational to electrostatic energy (growth rate of instability
in the bulk plasma) is comparable to the rate of transfer from
electrostatic to kinetic energy (damping rates on the fast particles)

Before we dwell on the LLCD led neutrino energization, let us 
recall various mechanisms in the literature that are likely to contribute to the observed flux of
PeV neutrinos (from all AGN):

1) In Atoyan \& Dermer (2001) the authors have assumed that in blazars, protons accelerate 
with the same power as electrons and are injected with the number spectrum $-2$. Under
these circumstances it has been shown that the pion photo production
process might lead to generation of very high energy neutrinos. The authors in Mannheim et al. (2001)
consider the limits of neutrino background based on the observationally evident 
intensity of extragalactic gamma-rays. An
analysis of  the photon spectrum (Cholis \& Hooper 2013) predicts 
AGN neutrinos with energies  even higher than detected
by the IceCube collaboration. Such a conclusion, however, is model
dependent

2) Two similar studies, investigating the production of UHE neutrinos, are
noteworthy: a) Stecker et al. (2013), the high energy neutrinos arise 
from AGN cores where particles are accelerated by shocks, and  b)
in Kalashev et al. (2015), the process of photopion production in
the Shakura-Sunyaev accretion disks is invoked to explain the
observed Ice Cube neutrino flux.

Our work, however, exploits a totally different process-  PeV neutrinos
are created by the centrifugally accelerated protons. Instead
of focusing on the accretion disk as a location where the neutrinos
might arise, we estimate the diffusive effect of all AGNs and show that the average fraction $0.003\%$ of
the bolometric luminosity of these objects is enough to produce the observed flux
of PeV neutrinos. To the best of our knowledge, this work
constitutes the first attempt of this kind. The present mechanism does not contradict aforementioned mechanisms of acceleration, but instead is an alternative process, which might take place in rotating magnetospheres.

This paper is organised as follows: in section 2 we review the LLCD mechanism, in section 3 we
consider the energies of neutrinos and the corresponding flux and in
section 4, we summarise our findings and results.

\section{Review of LLCD}

We will now summarise aspects of the LLCD
that are most relevant to the creation of UHE  protons, and,
consequently, the UHE neutrinos.

Acceleration of protons strongly depends on the angular velocity of
the black holes
\begin{equation}
\label{rotat} \Omega\approx\frac{a c^3}{GM}\approx
10^{-3}\frac{a}{M_8}rad/s,
\end{equation}
where $c$ is the speed of light, $M_8\equiv M/(10^8M_{\odot}$ is the
normalized mass of the supermassive black hole, $G\approx 6.67\times
10^{-8}$dyne-cm$^2$g$^{-2}$ is the gravitational constant and
$0<a\leq 1$ is a dimensionless parameter characterizing the spinning
rate of the black hole.

In the AGN (pulsar) magnetospheres, the magnetic field is strong
enough to justify the so called frozen-in assumption; the particles,
essentially, follow the co-rotating field lines. The magnetic field
in the AGN magnetosphere is estimated as (Osmanov 2008)
\begin{equation}
\label{mag} B\approx 87\times\left(\frac{L}{10^{43}erg\;
s^{-1}}\right)^{1/2}\times\frac{R_{lc}}{r}G,
\end{equation}
where $L$ is the luminosity, $R_{lc}\equiv c/\Omega$ is the light
cylinder radius and $r$ is the radial distance from the supermassive
black hole. One can straightforwardly check that the gyroradii of
electrons and protons, characterized with the Lorentz factors of the
order of $10^{4-6}$, are much less than the kinematic lengthscale,
$R_{lc}$. Therefore, the frozen-in condition will be maintained
almost during the whole course of motion.

But it turns out that the energies corresponding to the centrifugal
acceleration, are not high enough to account for the highly
energetic cosmic ray protons.  In Osmanov et al. (2014) and Osmanov (2008) we
 discussed  two major mechanisms  that limit the acceleration
process: the inverse Compton scattering, and the so called
"breakdown of the bead on the wire approximation" \cite{rm00,r11}.
It has been shown that the latter is much more efficient and limits
the maximum attainable Lorentz factor to
$$\gamma_{_{BBW}}\approx\frac{1}{c}\left(\frac{e^2L}{2m}\right)^{1/3}\approx$$
\begin{equation}
\label{gfb} \;\;\;\;\;\;\;\;\;\;\;\;\;\;\;\;\;\;\;\;\approx 3\times
10^{5}\left[\frac{m_p}{m}\times\frac{L}{10^{43}erg\;
s^{-1}}\right]^{1/3},
\end{equation}
where $e (m)$ is the electron charge (mass); the latter is
normalized on the proton mass, $m_p$. The $\gamma_{BBW}$ is not high
enough for the cosmic ray energy range that we seek to explain. One
obtains a similar result  for electrons.

\begin{figure}
  \centering
  \includegraphics[width=0.55\textwidth]{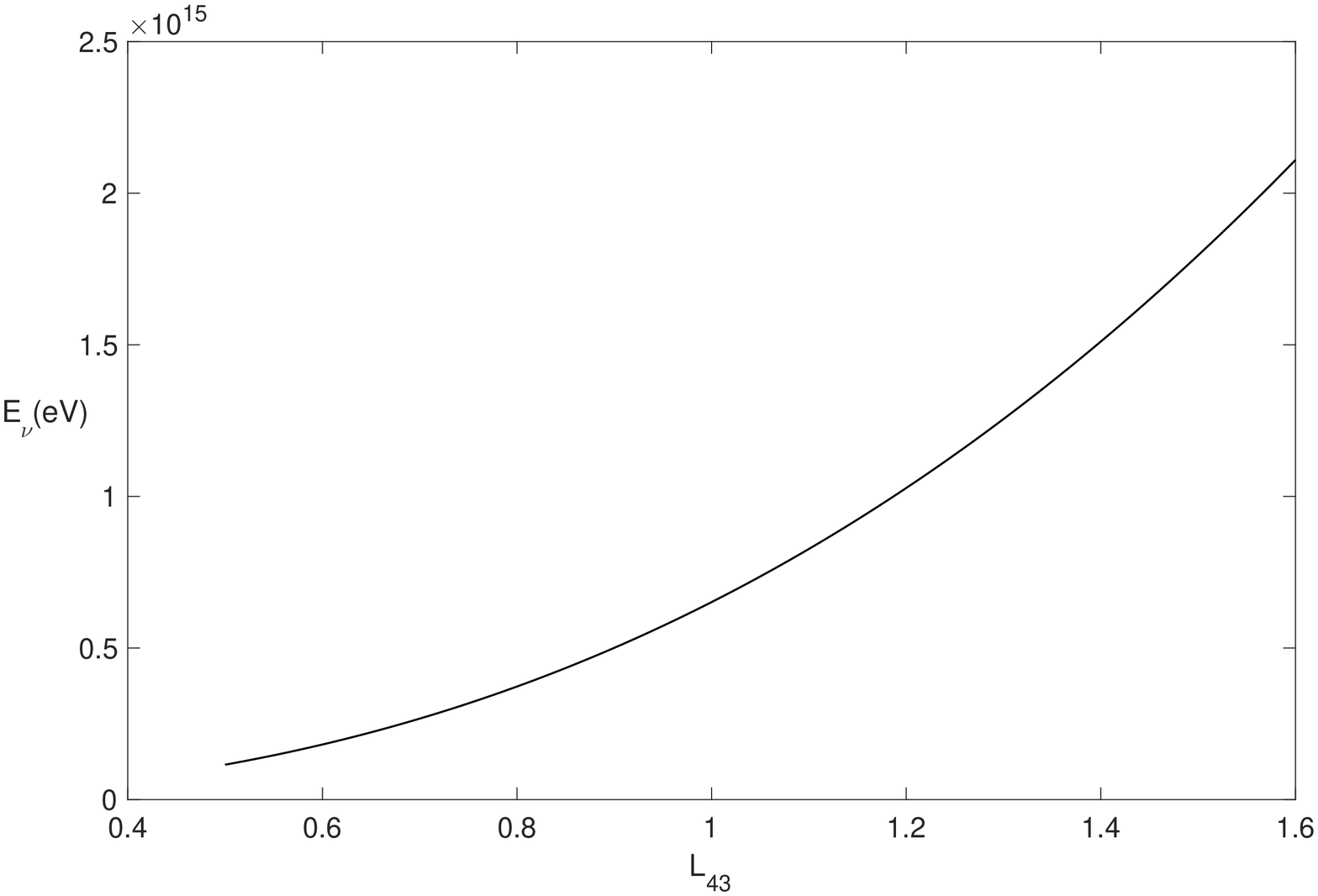}
  \caption{Dependence of energies of the produced neutrinos on the AGN luminosity normalised by $10^{43}$ergs s$^{-1}$. The set of parameters is: $\gamma_p = 100$, $g=0.001$ and $M_8 = 1$.}\label{fig1}
\end{figure}

Direct centrifugal acceleration, therefore, cannot boost particle
energies up to the very-high energy domain. The relativistic
centrifugal effects, however, do play an indirect role in additional
particle acceleration: it is this very time-dependent, differential
centrifugal acceleration that, first, transfers the rotational
energy into Langmuir waves  via a  parametric two stream instability
with a growth rate (Osmanov et al. 2014)
\begin{equation}
 \label{grow}
 \Gamma= \frac{\sqrt3}{2}\left (\frac{\omega_e {\omega_p}^2}{2}\right)^{\frac{1}{3}}
 {J_{b}(b)}^{\frac{2}{3}},
\end{equation}
where $J_{\mu}(x)$ denotes the Bessel function, $b=\omega_e/\Omega$,
$\omega_{e,p}\equiv\sqrt{4\pi e^2n_{e,p}/m_{e,p}\gamma_{e,p}^3}$,
$n_{e,p}$ and $\gamma_{e,p}$ are, respectively, the relativistic
plasma frequency, the number density and the Lorentz factor of the
corresponding specie (electrons or protons). By considering the
parameters $L\sim 10^{43}$erg$\;$ s$^{-1}$,
$\gamma_e\sim\gamma_{_{BBW,e}}$, ($\gamma_{_{BBW,e}}$ is the maximum
attainable Lorentz factor of electrons by direct centrifugal
acceleration) $\gamma_p\sim 10^2$, after taking into account the
equipartition distribution of energy among protons with different
energies, one can show that the instability time-scale is less than
the kinematic time-scale, $R_{lc}/c$, by three orders of magnitude. We assume that energy distribution of particles is continuous and thus, apart from the ultra-high energies, there are relativistic protons with relatively low Lorentz factors. As we have shown, the parametric instability for particles with $\gamma_p\sim 100$ is extremely high and only this particular class of protons can very efficiently contributes in the mentioned process. It is worth noting that despite the fact that Eq. (\ref{grow}) is derived in the linear approximation, the driving force - the centrifugal force - might be very large and consequently the energy pumped into the waves is huge even in the linear regime. On the other hand, as it has been discussed in detail by Osmanov et al. (2014) and Mahajan et al. (2013) the parametrically unstable Langmuir modes have superluminal phase velocities and since there are no particles with such speeds, the electrostatic waves will not Landau damp increasing their energy until the Langmuir collapse.

In the second stage, the growing energy of the Langmuir waves is
efficiently extracted by particles in a two stage process: the wave
energy is spatially concentrated through what is called the Langmuir
Collapse \cite{zakh,arcimovich}, and then it Landau damps on the
faster particles; the collapse does not develop in the inner
magnetosphere ($r<R_{lc}$, where the magnetic field is dominant
\cite{zev}. Outside this zone, the energy of the electrostatic field
evolves as \cite{arcimovich}
\begin{equation}
\label{E2} 
\;\;\;\;\;\;\;\;\;\;\;\;\;\;\;\;\;\;\;\;\;\;\;\;\;\;\;\;\;\frac{\mid E\mid^2}{8\pi}\approx \frac{\mid
E_0\mid^2}{8\pi}\left(\frac{t_0}{t_0-t}\right)^2
\end{equation}
where $E_0$ is the initial value of the electric field and $t_0$ is
the time of "complete" collapse. The energy density of the electric field
grows explosive, pulling protons from the caverns, and then
efficiently transferring energy from the electrostatic waves to the
particles via  Landau damping. As a result, the protons are
driven to UHE (Osmanov et al. 2014),
$$\epsilon_p(eV)\approx\frac{ne^2}{4\pi^2\lambda_D^3}\Delta r^5\approx$$
\begin{equation} \label{energy} 
\;\;\;\;\;\;\;\;\;\;\;\;\;\;\;\;\;\approx 1.14\times
10^{17}\times\left(\frac{g}{10^{-3}}\right)^3 \times
M_8^{-5/2}\times L_{43}^{5/2},
\end{equation}
where $\Delta r\approx R_{lc}/\left(2\gamma_p\right)$ is the width
of a narrow region close to the LC where the process of acceleration
takes place, $\lambda_D$ is the Debye Length, and
\begin{equation}
\label{n} 
n=\frac{L}{4\eta\pi m_pc^2\upsilon R_{lc}^2}\approx
6.3\times 10^6\times L_{43}\times R_{lc,8}^{-2} cm^{-3}
\end{equation}
is the approximate value of the proton number density outside of the
LC, $L_{43}\equiv L/(10^{43}erg\; s^{-1})$, $R_{lc,8}\approx
3M_8\times 10^{14}$cm is the light cylinder radius, and $g$ is the
initial dimensionless density perturbation. This estimate is done
for $\gamma_p = 100$. In the current paper we have assumed that only 10 per cent of the rest energy of accretion matter ($\eta = 0.1$) transforms to emission. 

It is clear that during the process of acceleration particles might potentially lose energy and the corresponding scenario has to be examined as well. This particular problem has been studied by Osmanov et al. (2014) and for detailed analysis a reader is referred to it. Here we only briefly outline the results of the mentioned work. The synchrotron mechanism of energy losses is so efficient that soon after the beginning of motion the protons lose their transverse momentum, transit to the ground Landau level and  the synchrotron process does not affect dynamics any more. For the inverse Compton mechanism it can be shown that the timescale of acceleration behaves linearly with the gained energy and consequently cannot compete with the acceleration timescale, which is a continuously decreasing function of $\epsilon_p$. The similar result is valid for the curvature radiation, which as it has been estimated in (Osmanov et al. 2014) can be significant only for energies exceeding $10^{17}$eV by many orders of magnitude.

\section{Neutrino energies and the flux}

In this section we address the significant questions concerning the energies of neutrinos and discuss the produced integral flux by taking into account AGNs having appropriate luminosities.

From Eq. (\ref{enu}) we obtain
\begin{equation}
\label{ep}
\epsilon_{p,17}\approx\frac{1}{4}\times\left(1+z\right)\times\frac{E_{\nu}}{PeV}.
\end{equation}
Therefore, for $z_{max} = 6$, one finds that neutrinos in the energy
interval $[1-2]$PeV will correspond to proton energies in the range
$\epsilon^{min}_{p,17}\approx 0.22$ - $\epsilon^{max}_{p,17}\approx
3.07$. One can further infer from
 Eq. (\ref{energy}) that if the bolometric luminosity of AGN were in the range
$L_{min}\approx 5.45\times 10^{42}$ ergs s$^{-1}$ - $L_{max}\approx
1.57\times 10^{43}$ ergs s$^{-1}$, protons of required energy could
be crated  via LLCD. In Fig. \ref{fig1} we show the behaviour of energies of neutrinos on AGN luminosities 
normalised by $10^{43}$ergs s$^{-1}$. The corresponding set of parameters is: $\gamma_p = 100$, $g=0.001$ and $M_8 = 1$. As it is clear from the plot $E_{\nu}(eV)$ is a continuously increasing function of the AGN luminosity, which is a natural result, because more luminous AGNs produce more energetic neutrinos.

Since all AGNs are supposed to be rotating supermassive black holes,
one can make a case that the observed UHE neutrino fluxes could,
at least partially, originate in their vicinity. IceCube
collaboration has announced that the neutrino flux in the energy
interval $100$TeV-1PeV is of the order of $10^{-8}$GeV
cm$^{-2}$s$^{-1}$sr$^{-1}$ \cite{pev2}. Can the proposed LLCD
mechanism lead to such neutrino fluxes? And if yes, what is
the fraction of the bolometric luminosity ($\mu$) that goes to PeV
neutrinos?

To answer these questions, let us estimate the neutrino flux
coming out of our theory. Since the cosmic ray kinetic luminosity, $L_{CR}$ (the total kinetic energy of protons 
"emitted" by the AGN per second),
cannot exceed the bolometric luminosity of AGN,  the flux provided
by this source takes the form \cite{weinb}

\begin{equation}
\label{sf} F_0(L) = \frac{\chi LR(t_1)^2}{4\pi R(t_0)^4r_1^2},
\end{equation}
 $R$ denotes the scale factor in the Robertson-Walker
metric, $t_1$ is time when the cosmic rays were emitted from the
source and $t_0$ is time when the cosmic rays reached a detector,
$r_1$ is the radial distance of a source at the time $t_1$ and
$\chi\equiv L_{CR}/L<1$.  If $\phi(t_1,L)$  measures the number of
AGNs (per unit of volume/ per unit  luminosity), the actual flux
of cosmic rays for the luminosity interval $[L; L+dL]$ and the
radial distance interval $[r_1; r_1+dr_1]$ is given by \cite{weinb}
\begin{equation}
\label{df} dF = F_0(L)dN = 4\pi
F_0(L)R(t_1)^2r_1^2\phi(t_1,L)|dt_1|\frac{dL}{\overline{L}},
\end{equation}
where $dN$ is the number of AGN for the aforementioned intervals and
$\overline{L}$ is the exponential luminosity cut-off.

By taking into account the relations $R(t_0)/R(t_1)=z+1$,
$\phi(t_1,L) = \phi(L)\left(1+z\right)^3$, $|dt_1| =
cH_0^{-1}\left(1+z\right)^{-5/2}dz$, where $H_0 = 67.8$km/s/Mpc is
the Hubble constant \cite{planck}, and $z$ is the redshift of the
object (we have used the value $q_0 = 1/2$ \cite{weinb}) Eq.
(\ref{df}) leads to the following cosmic ray flux
$$F =
\frac{2}{5H_0\overline{L}}\left[1-\left(1+z_{_{max}}\right)^{-5/2}\right]\times$$
\begin{equation}
\label{flux} 
\;\;\;\;\;\;\;\;\;\;\;\;\;\;\;\;\;\;\;\;\;\;\;\;\;\;\;\;\times
\int_{L_{min}}^{L_{max}}F_0(L)\phi(L)dL,
\end{equation}
$L_{min}$ ( $L_{max}$) denotes minimum (maximum) values of AGN
luminosities, and $z_{_{max}}$ is the redshift of the most distant
AGN. To our knowledge there is one at a redshift in excess of $5$.
This is the object described in Barger et al. (2002) with $z=5.186$, in the
Chandra Deep-Field. For the current work, we will simply use
$z_{max}=6$. To estimate the above integral, one needs the
luminosity function, $\phi(L)$, describing the distribution of
galaxies in general, and AGN in particular. Observationally, the
best fit to the luminosity function has the form
\begin{equation}
\label{lumfun}
\phi(L)=\phi_0\left(\frac{L}{\overline{L}}\right)^{\alpha}e^{-L/\overline{L}},
\end{equation}
where $\phi_{0}= 0.012$h$^3$Mpc$^{-3}$, $\alpha = -1.25$ defines the
slope of the power law and $\overline{L} = 1.32\times 10^{44}$erg
s$^{-1}$ \cite{longair}.

With this luminosity function, the neutrino diffusive flux comes
out to be
$$F_{\nu}^{theory} = \frac{2\mu c \phi_0\overline{L}}{5H_0}
\left[1-\left(1+z_{_{max}}\right)^{-5/2}\right]\times$$
\begin{equation}
\label{neflux}
\;\;\;\;\;\;\;\;\;\;\;\;\;\;\;\;\;\times\left[\Gamma\left(\alpha+2,\frac{L_{min}}{\overline{L}}\right)-
\Gamma\left(\alpha+2,\frac{L_{max}}{\overline{L}}\right)\right],
\end{equation}
where $\Gamma\left(t, x\right)$ is the incomplete Gamma function.

From observations, the best power law fit for the  neutrino flux
may be expressed  as \cite{pev2}
$$f(E_{\nu})\equiv
E_{\nu}^2\frac{dN_{\nu}}{dE_{\nu}}\approx$$
\begin{equation}
\label{Iceflux} \;\;\;\;\;\;\;\;\;\;\;\approx 1.5\times
10^{-8}\times\left(\frac{E_{\nu}}{0.1PeV}\right)^{-0.3}GeV\;
cm^{-2}s^{-1}sr^{-1}.
\end{equation}
The number of neutrinos carrying energy in the interval $E_{\nu};
E_{\nu}+dE_{\nu}$ is $dN_{\nu}=E_{\nu}^{-1}f(E_{\nu})dE_{\nu}$,
therefore, the total energy flux of neutrinos in the energy
interval $[1-2]$PeV may be written as
$$F_{\nu}^{obs} =
4\pi\int_1^2E_{\nu}^{-1}f(E_{\nu})dE_{\nu}\approx $$
\begin{equation}
\;\;\;\;\;\;\;\;\;\;\;\label{Iceflux1} \approx 9.47\times
10^{-11} ergs\;s^{-1} cm^{-2},
\end{equation}
where where the factor 
$4\pi$ takes into account the total solid angle.

Comparing equations (\ref{neflux}) and (\ref{Iceflux1}), we may
conclude that the theoretical predictions for the PeV neutrino flux
will be consistent with observations if a rather small fraction (
$\mu\approx 3.08\times 10^{-5}$) of the bolometric luminosity were
converted to power UHE neutrinos.

\section{Summary}

\begin{enumerate}

\item

We have presented, here, a theoretical pathway for generating high
energy neutrinos in the range $[1-2]$PeV. These neutrinos are
produced in hadronic reactions, in particular, by very energetic
protons.

\item
The first step in the theory, therefore, was to find an
astrophysical setting where ultra energy protons are produced. Such
a setting was recently explored: it was shown that in the rotating
magnetospheres of AGN, UHE protons ( ~$10^{17}$eV) could, indeed, be
created through LLCD-a two step process that converts the free
energy in differential rotation, first to the electrostatic energy
in Langmuir waves, and then to particle energy through landau
damping of Langmuir waves (Osmanov et al. 2014).

\item
Protons with this enormous energy could, then, create ultra high
energy neutrinos, recently observed by the IceCube collaboration. By comparison of theoretically 
derived expression of diffusive energy flux and the observed energy flux for the range $[1-2]$PeV
we have estimated the value of $\mu\approx 3.08\times 10^{-5}$ indicating that only a tiny fraction 
of the total bolometric luminosity is enough to explain the neutrino flux.

      \end{enumerate}

This work is a first attempt of this kind, where we considered the role of centrifugal mechanism of acceleration in generation of cosmic protons with energies enough to produce the PeV neutrinos and estimated the average fraction of the total luminosity of AGNs that is responsible for producing the PeV neutrinos.

\section*{Acknowledgments}

We would like to thank Dr. R. Shanidze for valuable comments. The research of ZO, GM and NC was supported by Shota Rustaveli National Science Foundation Grant NFR17-587, the work of ZO was partially supported
by Shota Rustaveli National Science Foundation Grant DI-2016-14 and the research of SM
was, in part, supported by USDOE Contract No.DE-- FG 03-96ER-54366.

\end{document}